\documentclass[12pt]{article}
\usepackage{a4wide}
\usepackage{amssymb}
\usepackage{graphicx}
\begin{document}
{\renewcommand{\thefootnote}{\fnsymbol{footnote}}
\begin{center}
{\LARGE  Algebraic properties of quantum reference frames: Does time fluctuate?}\\
\vspace{1.5em}
Martin Bojowald$^1$\footnote{e-mail address: {\tt bojowald@psu.edu}}
and Artur Tsobanjan$^2$\footnote{e-mail address: {\tt arturtsobanjan@kings.edu}}
\\
\vspace{0.5em}
$^1$ Institute for Gravitation and the Cosmos,
The Pennsylvania State
University,\\
104 Davey Lab, University Park, PA 16802, USA\\
\vspace{0.5em}
$^2$ King's College, 133 North River Street, Wilkes-Barre, PA 18711, USA
\vspace{1.5em}
\end{center}
}

\setcounter{footnote}{0}

\begin{abstract}
  Quantum reference frames are expected to differ from classical
  reference frames because they have to implement typical quantum features
  such as fluctuations and correlations. Here, we show that fluctuations and
  correlations of reference variables, in particular of time, are restricted
  by their very nature of being used for reference. Mathematically, this
  property is implemented by imposing constraints on the system to make sure
  that reference variables are not physical degrees of freedom.  These
  constraints not only relate physical degrees of freedom to reference
  variables in order to describe their behavior, they also restrict quantum
  fluctuations of reference variables and their correlations with system
  degrees of freedom. We introduce the notion of ``almost-positive'' states as
  a suitable mathematical method. An explicit application of their properties
  to examples of recent interest in quantum reference frames reveals
  previously unrecognized restrictions on possible frame-system
  interactions. While currently discussed clock models rely on assumptions
  that, as shown here, make them consistent as quantum reference frames,
  relaxing these assumptions will expose the models to new restrictions that
  appear to be rather strong.  Almost-positive states also shed some light on
  a recent debate about the consistency of relational quantum mechanics.
  \end{abstract}

\section{Introduction}

The unavoidable fact that measurements are made with measurement devices
implies that a complete description of the underlying process must use degrees
of freedom for both the measured and the measuring systems. In quantum
mechanics, both are represented by states or, rather, a single combined state,
and they may be expected to be subject to fluctuations, correlations, and
entanglement. The accessibility and consistency of relationships between a
measured system and a frame set up to describe the measuring device has given
rise to much recent work \cite{QuantumRef, QuantumRef1, QuantumRefSpin,
  QuantumRef2, QuantumRef3, QuantumRef4, QuantumRef5, QuantumRef7}. A common
method is the application of quantized constraints to impose the reference
relations between parts of the combined system. The purpose of the present
paper is to point out that constrained reference states produced by imposing
quantum constraints are distinct from ordinary entangled states in a precise
mathematical sense. The essential feature of constrained states is that they
contain redundant information, which is used in computations of probabilities
for outcomes of physical measurements, but is not itself directly measurable
or subject to dynamics. While we will focus here on temporal reference frames,
much of the discussion is general and applies whenever quantum constraints are
used.

Examples of redundant information abound in space-time physics (and are
therefore prominent in the endeavor of quantum gravity), the most important
one being the coordinates used to describe motion in space-time. In our
discussion, it will be important to distinguish between the values of
coordinates (time or space) and the physical devices used to measure them
(clocks or rulers, or their generalized versions). The former amount to
redundant information because they may well be chosen differently. It is
important to determine how one could transform between different coordinate
choices in order to compare different descriptions of the same physics, but
this task is distinct from the determination of physical properties of space
and time.  A clock or a ruler or some other spatial measurement device, by
contrast, is a physical instrument that, in a sensitive quantum measurement,
may well interact with the system in some way and influence the measurement
outcome. Quantum fluctuations or entanglement between clock and system may
then be relevant. However, a clock is not the same as time, and a ruler is not
the same as space. Therefore, their mathematical treatments and physical roles
may well be different from each other. A dedicated analysis is required in
order to determine whether time and space (as opposed to clocks and rulers)
should be subject to quantum properties such as fluctuations and entanglement.

In \emph{ordinary} quantum mechanics the state of a system is typically
described relative to some external classical reference frame using
``coordinates'' (or analogous references) that have no inherent physical
significance to the system being described. Changing the coordinate frame (for
example by rotating Cartesian axes) therefore leads to a physically equivalent
state of the system seen from a different coordinate
perspective. Schematically, we can represent this transformation as
\[
\sum_{c} \psi_a(c) |c \rangle_a  \rightarrow \sum_{c} \psi_b (c) |c \rangle_b  .
\]
Here $a$\ and $b$\ label classical reference systems which are not part of the
quantum degrees of freedom, $\psi_i(c)$\ are numerical coefficients, and
e.g. $|c\rangle_a$\ denotes the basis in which $c$\ is described in frame $a$
(which may or may not coincide with $|c\rangle_b$). Ultimately, however,
measurement references are made with respect to physical objects (clocks,
rulers, etc.) which are fundamentally quantum mechanical, and are therefore
subject to the usual quantum mechanical effects such as fluctuation,
discreteness, and quantum superposition. One can explore the quantum effects
of reference frames by incorporating them into the quantum mechanical
description. A measurement device $a$\ will start in some pre-measurement
state $| \phi_a \rangle$\ and the measurement of $c$\ relative to $a$\ can be
modeled by some interaction Hamiltonian which will induce a transformation on
the combined state of $a$\ and $c$ to an entangled state
$| \phi_a \rangle \otimes \left( \sum_c \psi_a (c) | c \rangle_a \right)
\rightarrow \sum_c \psi_a (c) |a (c) \rangle \otimes | c \rangle_a$, where
state $|a(c)\rangle$\ of the device correlates with state $|c \rangle_a$\ of
the measured system. Frame transformation therefore corresponds to linking a
pair of entangled states
\[
\left( \sum_c \psi_a (c) |a (c) \rangle \otimes | c \rangle_a \right) \otimes |\phi_b \rangle \rightarrow \left( \sum_c \psi_b (c) |b(c) \rangle \otimes | c \rangle_b \right) \otimes |\phi_a \rangle \ .
\]

The ongoing work on quantum reference frames deals with the quantum nature of
measurement references in a structurally different way (see for
example~\cite{QuantumRef1}). Here, in any given reference frame, one
constructs a more or less ordinary quantum mechanical description of the
observed system of interest as well as of all available reference frames other
than the one being used. Schematically, given quantum reference frames $a$ and
$b$, and using $c$\ to denote ``the rest'' of the quantum system being
modeled, the transformation from frame $a$\ to frame $b$\ has the form
\[
\sum_{b, c} \psi_a (b, c) |b \rangle \otimes|c \rangle  \rightarrow \sum_{a, c} \psi_b (a, c) |a \rangle \otimes|c \rangle  .
\]
In a given reference frame, the degrees of freedom corresponding to the frame
itself are not part of the quantum representation, however they appear in the
quantum description relative to a different reference frame.

Let us compare the schematic workings of a quantum reference frame to the two
uses of reference systems in ordinary quantum mechanics described earlier. It
is immediately obvious that quantum reference frames are in general
inequivalent to classical reference frames. Classical reference systems are
not dynamical within the quantum description, while, at least from the
perspective of quantum reference frame $a$, quantum reference frame $b$\ can
be quantum mechanically coupled to the observed system. At first,
incorporating quantum references as subsystems within an ordinary quantum
mechanical description appears as a promising way to simultaneously treat
quantum degrees of freedom of $a$, $b$, and $c$, and then to somehow project
to individual quantum reference frame descriptions. At the very least, this
would require additional couplings, since otherwise subsystem $a$\ will not
have a description of subsystem $b$, but only of subsystem $c$, etc. The
viability of embedding a quantum reference frame within a larger ordinary
quantum state description of all quantum degrees of freedom can be further
restricted if we use some of the details of how quantum reference frames are
constructed. Here we turn to some of the work on quantum reference frames that
implements quantum frame covariance using Dirac constraint quantization
\cite{QuantumRef2, QuantumRef3}, especially in the case of temporal references
\cite{QuantumRefSwitch, Switching, QuantumRef4}. The advantage of using quantum
constraints is that the process of establishing a quantum reference frame
proceeds via a perspective neutral framework of constrained states defined on
the algebra of operators that represent the degrees of freedom of all of the
reference frames at once.

Here we use a very general algebraic analysis of constrained quantum systems
developed in~\cite{AlgebraicTime} to explore the properties of temporal
quantum reference frames. In section~\ref{sec:Constraint} we describe a simple
model of a Hamiltonian constraint and briefly review its classical properties
as the regulator of temporal reference frames. In
section~\ref{sec:Entanglement} we review some general properties of
constrained quantum states and in section~\ref{sec:APstates} we introduce a
class of constrained states that have a relational interpretation. We argue
that constrained states that have an interpretation as (temporal) reference
states are quite different from correlated states of quantum mechanics in that
the former encode redundancy rather than correlation. We further argue that,
in light of this redundancy, a temporal quantum reference frame is not the
same as a physical clock. In section~\ref{sec:GaugeFlows} we discuss the
restrictions that the form of the Hamiltonian constraint places on the
viability of a given temporal quantum reference frame. In
section~\ref{sec:Clocks} we discuss the implication of these restrictions on
the viability of interacting physical clocks (such as the models as considered
in~\cite{QuantumRefLocal, QuantumRef6, QuantumRefEquiv}) as temporal reference
variables.

\section{Hamiltonian constraint and temporal reference frames} \label{sec:Constraint}

In canonical descriptions of space-time physics, redundancy of information is
mathematically incorporated through the use of constraints. One first
describes the properties of a system using degrees of freedom that contain
redundancies, and then imposes constraints to make sure that only physical
degrees of freedom remain. In quantum gravity, suitable constraints arise
automatically within the canonical formulation of general relativity, in which
general covariance is realized not manifestly by transformations on space-time
tensors but rather by restrictions on allowed configurations.

We will build our present discussion around the technically much simpler
example of time-reparameterization-invariant formulation of an ordinary
classical mechanical system. Starting with an $N$--component system with
degrees of freedom $q_i$, $p_i$, $i=1, 2 \ldots N$, and a Hamiltonian
$H(q_i, p_i;t)$, we extend the phase-space to include a canonical pair given
by time $q_0$\ and its canonically conjugate momentum $p_0$ (often referred to
as the energy of the ``clock'' $q_0$). Instead of a Hamiltonian, the redundant
dynamics is captured by the Hamiltonian constraint
\begin{equation}\label{eq:Cform}
C=p_0+H(q_i,p_i;q_0) \ ,
\end{equation}
where $i\neq0$\ and we have replaced parameter time $t$ in the Hamiltonian by
the new phase-space time variable $q_0$. Physically allowed states of the
system lie on the constraint surface $C=0$ where
$p_0=-H(q_i,p_i;q_0)$. Hamilton's equations generated by $C$\ through the
canonical Poisson bracket are
\begin{equation}\label{eq:q_0Evol}
  \frac{{\rm d}q_0}{{\rm d}\tau}= \{q_0, C\} = \frac{\partial C}{\partial p_0}=1 \quad , \quad  \frac{{\rm d}p_0}{{\rm d}\tau}= \{ p_0, C\} = - \frac{\partial C}{\partial q_0}=0 \ , 
\end{equation}
and similarly for $i\neq0$
\begin{equation} \label{eq:q_iEvol}
\frac{{\rm d}q_i}{{\rm d}\tau}=\{ q_i, C\}=\frac{\partial    H}{\partial p_i} \quad , \quad \frac{{\rm     d}p_i}{{\rm d}\tau}=\{ p_i, C\}=-\frac{\partial    H}{\partial q_i}\ .
\end{equation}
In this view, the time and mechanical degrees of freedom are restricted to the
constraint surface and evolve along orbits generated by $C$\ relative to an
external parameter $\tau$. However, this parameter keeping track of evolution
along the Hamiltonian flow of $C$\ is arbitrary. Rescaling the constraint by
any non-vanishing phase-space function $\mathcal{N}$, gives us an entirely
equivalent constraint $\tilde{C}=\mathcal{N}C$: it defines the same constraint
surface as $\tilde{C}=0$\ implies $C=0$, and the flow that it produces on the
constraint surface reads
\[
 \frac{{\rm d}f}{{\rm d}\tilde{\tau}}= \{f, \tilde{C}\} =\{f, \mathcal{N}\}C+\{f, C\} \mathcal{N} \approx \{f, C\} \mathcal{N} = \mathcal{N} \frac{{\rm d}f}{{\rm d}\tau} \ ,
 \]
 where ``$\approx$'' denotes equality on the constraint surface and $f$\ is
 any phase space function. Evidently,
 ${\rm d}/{\rm d}\tilde{\tau}= \mathcal{N} {\rm d}/{\rm
     d}\tau$\ and the new flow is simply a reparameterized version of the
 original one. Because of this freedom of parameterization, it may be more
 fruitful to think of constrained evolution as phase-space degrees of freedom
 evolving relative to each other rather than relative to the arbitrary flow
 parameter $\tau$.

 This constrained system can be easily reduced to its unextended form in two
 steps. First we eliminate $p_0$\ using the constraint
 $p_0=-H(q_i,p_i;q_0)$. Second, we note that, according to~(\ref{eq:q_0Evol}),
 along the orbits generated by the constraint function
 $q_0(\tau)=const.+\tau$. We can therefore simply replace $q_0$\ in the
 Hamiltonian with a parameter time $t=const.+\tau$\ and
 ${\rm d}/{\rm d}\tau = {\rm d}/{\rm d}t$. This reduces the system to just the
 non-time degrees of freedom $q_i$, $p_i$, $i=1, 2 \ldots N$, which evolve
 relative to an external parameter time $t$\ along orbits of the Hamiltonian
 $H(q_i, p_i;t)$.

 An important feature of the constrained formulation of such a system is that
 reduction can be performed relative to any function on the extended phase
 space that has a non-vanishing Poisson bracket with $C$\ and therefore
 uniformly increases or decreases along its flow. Relative to a general time
 function and its conjugate momentum, the constraint does not have the simple
 form of equation~(\ref{eq:Cform}) and the analysis is geometrically more
 subtle. Nevertheless, in a classical treatment, for each suitable time, one
 gets a reduced unconstrained mechanical system with a Hamiltonian generating
 the dynamics relative to a time parameter. Each such distinct reduction can
 be interpreted as selection of the particular temporal reference frame
 associated with the corresponding time function. While the end results of
 reductions relative to different time functions will, in general, look quite
 different, they will nevertheless be linked via the ``timeless'' constrained
 framework: on the extended phase space different parameterizations of the
 flow of $C$\ on the constraint surface can be linked by following the flow
 itself with a suitable local rescaling $\mathcal{M}C$, where $\mathcal{M}$\
 is some (possibly vanishing) function of $q_0$, $p_0$, $q_i$, $p_i$.

 As in this simple example, constraints in canonical systems control
 redundancy in two ways: they constrain the extended phase space ($C=0$) and
 they generate gauge transformations (here along the parameter $\tau$) that
 relate variables on the constraint surface (after solving $C=0$) to flow
 parameters that transform some of the redundant information. In the given
 example, and in related but technically more involved examples from general
 relativity, the reduced system, where the leftover variables evolve relative
 to an external parameter, and the constrained system, where phase space
 degrees of freedom linked by redundancy evolve relative to each other, are
 clearly just two equivalent viewpoints of a given system. But the equivalence
 works only if we apply a suitable treatment to the constraint on the extended
 phase space. A relational interpretation of dynamics does not inherently
 necessitate a Hamiltonian constraint. For example, the Hamiltonian of the
 original $N$--component mechanical system generates time-evolution
 trajectories on the original phase space which (at least locally) can be
 interpreted as $(2N-1)$\ canonical variables evolving relative to the one
 remaining variable. In this situation, however, there is no redundancy in the
 description: all points of the original phase space represent allowed
 configurations (there is no constraint surface) that are physically distinct
 (there is no gauge flow).

At the quantum level, a constraint $C=0$ is implemented by the requirement
that a corresponding constraint operator, $\hat{C}$, annihilate all admissible
states. If a Hilbert-space representation is used for the quantum system, the
condition reads
\begin{equation} \label{Cpsi}
  \hat{C}|\psi\rangle=0
\end{equation}
for all admissible states $|\psi\rangle$. In the following section, we use generalized algebraic states to highlight important differences between states that satisfy the quantum constraint~(\ref{Cpsi}) and ordinary quantum states that contain correlations between subsystems.

\section{Redundancy versus correlation in the quantum setting}\label{sec:Entanglement}

Algebraically, a state is defined as a linear functional $\omega$ from the
operator algebra ${\cal A}$ of the system to the complex numbers. The
functional is required to be positive if it refers to observable information,
where positivity is defined by $\omega(\hat{A}^*\hat{A})\geq 0$ for all
$\hat{A}\in{\cal A}$, using a $*$-relation on the algebra that corresponds to
adjointness when represented on a Hilbert space. Such a state can be thought
of as an expectation-value functional that assigns the expectation value
$\omega(\hat{A})$ to an operator $\hat{A}$, just like
$\langle\psi|\hat{A}|\psi\rangle$ or ${\rm tr}(\rho\hat{A})$ in a
representation. It can be shown, see for instance \cite{LocalQuant}, that
every positive linear functional obeys the Cauchy--Schwarz inequality
\begin{equation}
  \omega(\hat{A}^*\hat{A})\omega(\hat{B}^*\hat{B})\geq
  |\omega(\hat{A}^*\hat{B})|^2
\end{equation}
for all $\hat{A},\hat{B}\in{\cal A}$, from which uncertainty relations follow
in the textbook manner. If the algebra contains a unit element,
$\mathbb{I}\in{\cal A}$, a positive state is real:
$\omega(\hat{A}^*)=\overline{\omega(\hat{A})}$ and therefore
$\omega(\hat{A})\in{\mathbb R}$ if $\hat{A}=\hat{A}^*$. A state is a minimal requirement
for a meaningful description of observable information, given by numbers
rather than operators, but it need not be constructed via a Hilbert-space representation
or wave functions.

Suppose $\mathcal{A}$\ is the operator algebra corresponding to an
unconstrained $N$--component quantum system, generated by
$\hat{q}_i, \hat{p}_i$, with $i=0, 1, 2, \ldots$, subject to the usual
canonical commutation relations
$[\hat{q}_i, \hat{p}_j] = i\hbar\, \delta_{ij}\mathbb{I}$, and suppose
$\omega$\ is an algebraic state in which, say, configurations of subsystems
labeled $0$\ and $1$\ are entangled. Then we make the following observations.
\begin{itemize}
\item Such a state makes probabilistic predictions for measurements of all
  observables of the system. Therefore it must be positive on the entire
  algebra $\mathcal{A}$\ as discussed above.
\item Entanglement requires that correlations between the observables
  $\hat{q}_0$\ and $\hat{q}_1$\ are non zero:
  $\Delta_{q_0q_1} =
  \frac{1}{2}\omega(\hat{q}_0\hat{q}_1+\hat{q}_1\hat{q}_0)-\omega(\hat{q}_0)\omega(\hat{q}_1)\neq0$,
  and similarly for other correlations involving powers of $\hat{q}_0$\ and
  $\hat{q}_1$.
\item Such a correlated state is special: since $[\hat{q}_0, \hat{q}_1] =0$,
  positivity of $\omega$\ places no bounds on the correlations between the
  configurations of the two subsystems. While there may be dynamical or
  symmetry considerations that strongly select for correlated states, the
  system possesses physically valid states in which the two subsystems are
  completely uncorrelated.
\item Even though they are correlated in an entangled state, both $\hat{q}_0$\
  and $\hat{q}_1$\ are individually observable. The state makes probabilistic
  predictions for measuring either of them individually without the other.
\end{itemize}

Now let us contrast this with an algebraic state on a constrained system. For simplicity, we will use the classical constraint of equation~(\ref{eq:Cform}) so that the quantized constraint is
\begin{equation}\label{eq:quantumC}
\hat{C}=\hat{p}_0+H(\hat{q}_i,\hat{p}_i;\hat{q}_0) \ ,
\end{equation}
with $i\neq 0$, where some suitable operator ordering has been chosen for the operator-valued function $H(\hat{q}_i,\hat{p}_i;\hat{q}_0)=H(\hat{q}_i,\hat{p}_i;\hat{q}_0)^*$. We enforce the quantum constraint~(\ref{Cpsi}) on an algebraic state by demanding
\begin{equation} \label{AC}
  \omega(\hat{A}\hat{C})=0\quad \mbox{for all}\quad \hat{A}\in{\cal A}\,.
\end{equation}
It is clear that the above conditions are satisfied by an ordinary null-eigenstate of the constraint operator, where expectation values can be constructed via the Hilbert-space inner product as usual $\langle \psi | \hat{A}\hat{C} | \psi \rangle$. However, a constraint operator of the form considered here will generically have zero in the continuous part of the spectrum. In a representation of algebra $\mathcal{A}$\ as acting on a Hilbert space $\mathcal{H}$ the null eigenstates of $\hat{C}$\ will belong to the space of linear functionals on $\mathcal{H}$ (the dual space $\mathcal{H}^*$). There will then be no default prescription for taking expectation values of operators relative to such states. Our generalized algebraic states, however, are not tethered to any particular Hilbert space representation of the algebra $\mathcal{A}$\ and there is no immediate obstruction for assigning numerical values to all operators in $\mathcal{A}$.

In contrast to an ordinary quantum state (entangled or not), \emph{an algebraic state that solves constraint~(\ref{eq:quantumC}) cannot be positive on the entire algebra $\mathcal{A}$}: we have $[\hat{q}_0,\hat{C}]=i\hbar \mathbb{I}$, and therefore
\begin{equation}
 \frac{1}{2} \omega(\hat{C}\hat{q}_0+\hat{q}_0 \hat{C}) = \omega (\hat{q}_0 \hat{C}) + \frac{1}{2}\omega([\hat{C},\hat{q}_0])= -\frac{1}{2}i\hbar \mathbb{I}
\end{equation}
using (\ref{AC}).  The result is imaginary, even though
$(\hat{C}\hat{q}_0+\hat{q}_0\hat{C})^*=(\hat{C}\hat{q}_0+\hat{q}_0\hat{C})$ as long as $\hat{C}^*=\hat{C}$ and $\hat{q}_0^*= \hat{q}_0$, making the corresponding state non-positive. 

While constrained quantum states cannot be positive on the entire algebra, they are still subject to certain positivity conditions. In a quantum system subject to a constraint $\hat{C}$, a notion of observables
is given by Dirac observables, defined as operators $\hat{O}\in{\cal A}$ such
that $[\hat{O},\hat{C}]=0$. Dirac observables therefore form a subalgebra $\mathcal{A}_{\rm obs}$\ of
${\cal A}$ given by the commutant of $\hat{C}$. Their action preserves the space of solutions of $\hat{C}$\ in any Hilbert space representation since $\hat{C}|\psi\rangle=0$ and $[\hat{O},\hat{C}]=0$ implies that
$\hat{C}\hat{O}|\psi\rangle=\hat{O}\hat{C}|\psi\rangle=0$. Furthermore, Dirac observables are invariant under the infinitesimal gauge flow generated
by $\hat{C}$ (which exponentiates to the unitary flow $\exp(i\tau\hat{C})$)---they are the quantum counterparts of extended phase space functions that are invariant along the flow generated by the constraint. Classically, such invariant functions are mapped to constants of motion during reduction of the constrained system relative to any valid time function. It therefore makes sense to insist that, in addition to condition~(\ref{AC}), physical algebraic states of a constrained system are positive on $\mathcal{A}_{\rm obs}$. Indeed, since Dirac observables commute with $\hat{C}$, condition~(\ref{AC}) does not in any way restrict positivity of the state on $\mathcal{A}_{\rm obs}$. Which brings us to a second contrasting point.

\emph{Constrained quantum states are not inherently entangled.} Suppose
$\hat{O}_1, \hat{O}_2 \in \mathcal{A}_{\rm obs}$\ are real observables,
$\hat{O}_i^*=\hat{O}_i$, correspond to different Dirac observable subsystems
$[\hat{O}_1, \hat{O}_2]=0$, and are not proportional to the constraint (so
that condition~(\ref{AC}) does not demand $\omega(\hat{A}\hat{O}_i)=0$\ for
all $\hat{A}\in \mathcal{A}$). In this case neither the constraint condition
nor positivity of states on $\mathcal{A}_{\rm obs}$\ place any restriction on
the value of subsystem correlations such as
$\Delta_{O_1O_2}=\frac{1}{2}\omega( \hat{O}_1 \hat{O}_2 + \hat{O}_2 \hat{O}_1
) - \omega( \hat{O}_1)\omega( \hat{O}_2)$\ which may well be small or zero.

Finally, \emph{constrained quantum states are not special states of the constrained system}. Without additional constructions, they are in fact the \emph{only} states on the full algebra $\mathcal{A}$\ that have a meaningful physical interpretation consistent with enforcing the quantum constraint.

\section{Almost-positive states} \label{sec:APstates}

Much of the immediately preceding discussion references Dirac observables. However, except for simple examples, it is usually hard to construct a complete set of Dirac observables, and, in fact, in a general situation, such a complete set may not exist \cite{DiracChaos,DiracChaos2}. While a constrained quantum state \emph{has to be} positive on $\mathcal{A}_{\rm obs}$\ and \emph{cannot be} positive on the entire algebra $\mathcal{A}$, there is a way to consistently extend positivity to additional subalgebras of $\mathcal{A}$\ that are more readily available. In~\cite{AlgebraicTime} we define a subclass of constrained states, which we christened {\em
  almost-positive states}, that are positive on a subalgebra associated with a reference observable and possess a relational interpretation. In this section we review some of the properties of these states and discuss additional ways in which they differ from ordinary quantum states that carry subsystem entanglement.

Any operator $\hat{Z} \in \mathcal{A}$\ that would correspond to a measurement in the absence of a constraint, so that $\hat{Z}^*=\hat{Z}$, can potentially serve as reference for constrained states. In a straightforward analogy with ordinary quantum mechanics, an observable $\hat{A} \in \mathcal{A}$\ can be determined simultaneously with $\hat{Z}$\ if $[\hat{A}, \hat{Z} ] =0$, and so without modifying the algebra structure, $\hat{Z}$\ can serve as reference for measuring all such observables. Analogously to $\mathcal{A}_{\rm obs}$, the commutant of $\hat{Z}$\ is a subalgebra, which we will denote by $\mathcal{A}_Z\subset \mathcal{A}$. The new definitions of \cite{AlgebraicTime} are based on the observation that
the commutant of a reference operator $\mathcal{A}_Z$ can, with some additional
constructions, replace the hard-to-obtain algebra of Dirac observables $\mathcal{A}_{\rm obs}$. Since
$\hat{Z}$ is usually a simpler operator than $\hat{C}$ because it does not
refer to the interacting dynamics, it is much easier to construct the
commutant of $\hat{Z}$ than the commutant of $\hat{C}$. In many cases,
$\hat{Z}$ may be one of the basic canonical operators such as $\hat{q}_0$ in our example, in which case $\mathcal{A}_Z$\ is simply the span of all basic
operators other than the conjugate momentum of $\hat{Z}$.

Just like in the classical example discussed in section~\ref{sec:Constraint},  the reference variable needs to characterize the flow generated by the constraint. It must therefore vary along this flow, so that $[\hat{Z},\hat{C}]\not=0$. Here we will require that $[\hat{Z},\hat{C}]=i\hbar \mathbb{I}$ in order to mimic the
energy-time relationship of our example, though this condition can be made somewhat more general. We immediately obtain the
result that the infinitesimal gauge flow of $\hat{C}$ preserves the commutant
of $\hat{Z}$, mimicking the corresponding property of Dirac observables:
According to the Jacobi identity, we
have
\begin{equation}
  [[\hat{A},\hat{C}],\hat{Z}]=[[\hat{Z},\hat{C}],\hat{A}]+[[\hat{A},\hat{Z}],\hat{C}]=0
\end{equation}
because $[\hat{Z},\hat{C}]$ is proportional to the identity operator, and
$[\hat{A},\hat{Z}]=0$ for $\hat{A}$ in the commutant of $\hat{Z}$. For any
$\hat{A}$ in  $\mathcal{A}_Z$, therefore, $[\hat{A},\hat{C}]$ is also in  $\mathcal{A}_Z$.

Through the condition in equation~(\ref{AC}) and positivity on
$\mathcal{A}_{\rm obs}$, we are already demanding that constrained states give
numerical probabilistic predictions for measurements of all Dirac observables
such that the constraint is identically zero. We now want to find a subset of
constrained states that, in addition, can also be used to assign probabilistic
predictions to all observables in $\mathcal{A}_Z$\ and correspond to a fixed
configuration of the reference observable $\hat{Z}$. We summarize these
conditions in our definition of almost-positive states, which are states
$\omega$ such that
\begin{enumerate}
\item they solve the constraint, $\omega(\hat{A}\hat{C})=0$ for all
  $\hat{A}\in{\cal A}$;
\item they parameterize $\hat{Z}$ as a reference variable,
  $\omega(\hat{Z}\hat{A})=\omega(\hat{Z})\omega(\hat{A})$ for all
  $\hat{A}\in{\cal A}$; and \label{def:AP2}
\item they are positive on the commutant of $\hat{Z}$,
  $\omega(\hat{A}^*\hat{A})\geq 0$ for all $\hat{A}\in {\cal A}_Z$.
\end{enumerate}
Note that parameterization of $\hat{Z}$\ in condition~\ref{def:AP2} above is
required only in the specified ordering where $\hat{Z}$ appears on the
left. The ordering is important for consistency with the commutation relations
involving $\hat{Z}$. For example, when computing the value of the commutator
$\omega([\hat{Z},\hat{E}])=\omega(\hat{Z}\hat{E}) - \omega(\hat{E}\hat{Z})$,
where $\hat{E}$\ denotes the conjugate momentum of $\hat{Z}$, we have to
re-order the product in the second term before we can apply the
parameterization condition; otherwise we would obtain the inconsistent result
$0=\omega(\hat{Z})\omega(\hat{E})-\omega(\hat{E})\omega(\hat{Z})=i\hbar$. For
$\hat{A}$ in the commutant of $\hat{Z}$, of course, re-ordering does not
change the expression.

One should be worried that none of the above conditions explicitly mention
positivity on $\mathcal{A}_{\rm obs}$. However, as we show
in~\cite{AlgebraicTime}, under some additional algebraic conditions on
$\hat{C}$, $\hat{Z}$\ and $\mathcal{A}$, which are satisfied by constraints of
the form~(\ref{eq:quantumC}), the above three conditions automatically imply
positivity of $\omega$\ on $\mathcal{A}_{\rm obs}$, without the need to
explicitly construct this algebra. The theorems in \cite{AlgebraicTime}
further show that the flow generated by the constraint preserves
almost-positivity and linearly evolves $\omega(\hat{Z})$. (This last statement
is somewhat obvious given that $[\hat{Z}, \hat{C}] = i\hbar \mathbb{I}$.) An
almost positive state $\omega$ can therefore be interpreted as a quantum state
on $\mathcal{A}_Z$\ at a fixed value of time $t_Z=\omega(\hat{Z})$\ and it can
be evolved to a different time using $\hat{C}$, such that reality and the
Cauchy--Schwarz inequality are preserved. (This is the algebraic analogue of
unitarity.)

In fact, almost-positive states provide a consistent embedding of the states
of a reduced quantum system within the states of its parent constrained
quantum system. (Almost-positive states are therefore more powerful than the
usual distinction between kinematical and physical Hilbert spaces.) Let us
illustrate this with the concrete example of the constraint of
equation~(\ref{eq:quantumC}) with $\hat{Z}=\hat{q}_0$. Here the degrees of
freedom of the reduced system are generated by $\hat{q}_i$\ and $\hat{p}_i$\
with $i\neq 0$. A solution of the reduced dynamics assigns a positive state on
this algebra for each value of parameter time $t$. This solution can be
linearly extended to a state on the commutant of $\hat{q}_0$, which is
generated by $\hat{q}_0$\ in addition to $\hat{q}_i$\ and $\hat{p}_i$\ with
$i\neq 0$, by setting $\omega(\hat{q}_0)=t$\ and using
condition~\ref{def:AP2}: since $t$\ is real the extension will remain positive
on the commutant of $\hat{q}_0$.\footnote{Conversely, enforcing positivity
  on the whole commutant of $\hat{Z}$, including $\hat{Z}$ itself,
  guarantees reality of the reference variable $\omega(\hat{Z})$.} The
constraint equation~(\ref{AC}) can then be used to further extend the state to
all of $\mathcal{A}$\ by setting
$\omega(\hat{A}\hat{p}_0)=-\omega\left( \hat{A}
  H(\hat{q}_i,\hat{p}_i;\hat{q}_0)\right)$. For each positive state of the
reduced system there is therefore a functional on the full algebra of the
parent constrained quantum system, which as we already discussed, cannot be
completely positive. Almost-positivity gives the specific ways in which
positivity must be relaxed on the constraint operator and the chosen reference
variable.

The parameterization condition of almost-positive states has an immediate
implication: the reference variable $\hat{Z}$\ is not a physical degree of
freedom but a parameter. If we apply the condition to $\hat{A}=\hat{Z}$, we
obtain $\omega(\hat{Z}^2)=\omega(\hat{Z})^2$ and therefore $\Delta Z=0$ for
quantum fluctuations of the reference variable. For any $\hat{A}$ in the
commutant of $\hat{Z}$ we obtain that the quantum correlations
\begin{equation}
  \Delta_{ZA}=\frac{1}{2}\omega(\hat{Z}\hat{A}+\hat{A}\hat{Z})-\omega(\hat{Z})\omega(\hat{A})=0
\end{equation}
vanish. A reference variable therefore does not fluctuate, and it cannot be
entangled with system degrees of freedom. It is correlated with its own
momentum, but not in a real way, since positivity does not extend to $\hat{E}$:
\begin{equation} \label{CZE}
  \Delta_{ZE}=\frac{1}{2}\omega(\hat{Z}\hat{E}+\hat{E}\hat{Z})-\omega(\hat{Z})\omega(\hat{E})=
  \frac{1}{2}\omega([\hat{E},\hat{Z}])+
  \omega(\hat{Z}\hat{E})-\omega(\hat{Z})\omega(\hat{E})= -\frac{1}{2}i\hbar\,.
\end{equation}
This result underlines the non-physical nature of the reference degrees of
freedom.\footnote{Interestingly, even though $Z$ does not fluctuate, the uncertainty
relation of the pair $(\hat{Z},\hat{E})$ is formally satisfied (and always saturated): $
  (\Delta Z)^2(\Delta E)^2- \Delta_{ZE}^2=\hbar^2/4\,. $}
If $\hat{Z}$ is the
quantum analog of a relational time, it is not a physical degree of freedom
but rather a parameter used to characterize redundant information. In our classical
example of section~\ref{sec:Constraint}, for instance, $q_0$ was simply replaced by the evolution parameter $t$ once the constraint has been solved and its gauge flow eliminated. Similarly, we expect a quantum reference variable to be replaced by a parameter (a number rather than an operator) once the constraints are solved. 

Viewed from this algebraic perspective, the reduced theory is not a description of co-evolution of a physical clock $\hat{Z}$\ in relation to observables compatible with $\hat{Z}$. Instead, it is a theory of reduced degrees of freedom unitarily evolving relative to a parameter time $t_Z$\ that, a priori does not possess a physical clock.

\section{Switching temporal reference frames}\label{sec:GaugeFlows}

The preceding section introduced almost positive states as a way to embed a
reduced quantum system within its parent constrained system with a particular
focus on the situation where the constraint governs dynamics. Similar to the
classical situation, the quantum constraint can, in general be reduced
relative to multiple internal times (though the situation is quite restrictive
as we shall see below). The simplest example of such a situation would be the
special case of constraint in equation~(\ref{eq:quantumC}) where one other
component subsystem (or more) behaves in the same way as $\hat{q}_0$
\begin{equation} \label{eq:CwClock}
\hat{C}=\hat{p}_0+\hat{p}_1+H(\hat{q}_i,\hat{p}_i;\hat{q}_0, \hat{q}_1) \ , \ \ i\geq2 \ .
\end{equation}
Here both $\hat{q}_0$\ and $\hat{q}_1$\ can be used as references and define almost-positive states. It is easy to see that almost-positive states with respect to $\hat{q}_0$\ will not be almost positive with respect to $\hat{q}_1$. From equation~(\ref{CZE}) we see that for any almost-positive state of $\hat{q}_0$\ we have $\Delta_{q_0p_0} = -i\hbar/2$. On the other hand, since $[\hat{q}_0, \hat{q}_1] = [\hat{p}_0, \hat{q}_1] = 0$, both $\hat{q}_0$\ and $\hat{p}_0$\ are in $\mathcal{A}_{q_1}$, so that an almost-positive state of $\hat{q}_1$\ is positive on their products, which requires $\Delta_{q_0p_0} \in \mathbb{R}$. 

In general, mapping the description of physics relative to one time variable
$\hat{Z}_a$\ to the description of the same physics relative to another time
variable $\hat{Z}_b$\ involves creating an identification between the two
corresponding sets of almost-positive states. Structurally, this is in a good
agreement with the general form of quantum reference frame transformations
discussed in the introduction. However, one might wonder why there would be
any such equivalence between algebraic constrained states considered here. In
the case where a complete set of Dirac observables is available (for an
$N$-component constrained canonical system that would mean that $2(N-1)$\
independent invariant operators have been constructed) the answer is
straightforward: a pair of constrained states are physically equivalent if
they assign the same values to $\mathcal{A}_{\rm obs}$. In the more general
situation considered in~\cite{AlgebraicTime} one looks at infinitesimal
properties of the flows on algebraic states that are generated by placing the
constraint operator on the left (as in $\omega(\hat{C}\hat{A})$), which are
guaranteed to preserve the value of Dirac observables. Our analysis
demonstrates that, once almost-positivity is imposed, the only equivalence
relation left is time evolution generated by $\hat{C}$, so at a fixed time
$t_Z=\omega_1(\hat{Z})=\omega_2(\hat{Z})$, different almost positive states
$\omega_1 \neq \omega_2$\ are physically distinct.

Since the algebraic analysis of~\cite{AlgebraicTime} focuses on infinitesimal
properties of physical equivalence relations, it does not provide a
straightforward method for constructing finite transformations between
temporal reference frames. Instead, it can be used to shed light on which
frames are viable. For a time variable that is part of a canonical subsystem,
like $\hat{q}_0$\ in our main example, our results show that (locally)
complete reduction requires the constraint to have the form given in
equation~(\ref{eq:quantumC}) or to factorize
$\hat{C}=\hat{\mathcal{N}} \hat{C}_{H}$, so that the right factor $\hat{C}_H$\
is of this form. In addition, adjointness relations $\hat{C}^*=\hat{C}$\ and
$\hat{C}_H^*=\hat{C}_H$\ severely restrict possible forms of the left factor
$\hat{\mathcal{N}}$. An example relevant to models considered
in~\cite{QuantumRef, QuantumRefLocal, QuantumRef6, QuantumRefEquiv} is a $\hat{C}$\
linear in $\hat{p}_0$. In this case
$\hat{\mathcal{N}} = \frac{1}{i\hbar}[\hat{q}_0, \hat{C}]$, and our results
require $\hat{\mathcal{N}}=\hat{\mathcal{N}}^*$,
$[\hat{\mathcal{N}}, \hat{C}_H]=0=[\hat{\mathcal{N}}, \hat{q}_0]$. For
example, this precludes the exact reduction of
\[
\hat{C} = \hat{q}_1 \hat{p}_0 + \frac{1}{2}\left(\hat{q}_1 \hat{p}_1 +\hat{p}_1 \hat{q}_1 \right) \ .
\]
Even though $\hat{C}$\ is hermitian, when we attempt to factorize we get
\[
\hat{C} = \hat{q}_1 \left( \hat{p}_0 + \hat{p}_1 -\frac{1}{2} i\hbar\, \widehat{\frac{1}{q}_1} \right) \ .
\]
where $\hat{\mathcal{N}} = \hat{q}_1=\hat{\mathcal{N}}^*$ and
$\hat{C}_H=\hat{p}_0 + \hat{p}_1 -\frac{1}{2} i\hbar\, \widehat{\frac{1}{q}_1}
\neq \hat{C}_H^*$. In such a situation we conclude that $\hat{q}_0$\ cannot be
used to exactly reduce the constraint by providing a temporal reference. This
still leaves the door open for approximate reduction under additional
conditions satisfied by states. Examples of current interest are shown in the
following section and in the appendix.

\section{Modeling time-keeping devices} \label{sec:Clocks}

We have so far not addressed an important question that naturally arises in
our analysis: Within the reduced theory, if the reference variable $\hat{Z}$\
is not an observable, then how could one measure time? This situation is not
that different from ordinary quantum mechanics where the evolution equation
for a system is specified relative to a parameter time. The studied system
itself does not by default possess a clock, but one can always add a subsystem
the measurement of which will correlate with (changes in) the value of
time. In the case of a constraint, where its right factor $\hat{C}_H$\ is
reduced by using $\hat{Z}$\ as the time reference (as in the previous
section), a physical clock must first and foremost be some observable of the
reduced system $\hat{U}=\hat{U}^* \in \mathcal{A}_Z$\ that is independent of
$\hat{Z}$. Furthermore, the clock must evolve along in time
$[\hat{U}, \hat{C}_H ] \neq 0$, with an \emph{ideal} clock evolving at a
constant rate so that $[\hat{U}, \hat{C}_H ] = i\hbar \mathbb{I}$\ up to a
real constant factor. A simple model for this situation is provided by the
constraint in equation~(\ref{eq:CwClock}), where $\hat{q}_0$\ can play the
role of reference time $\hat{Z}$\ and $\hat{q}_1$\ the role of the ideal
physical clock $\hat{U}$. Indeed, the roles can also be reversed: an ideal
clock is also a valid time reference variable. However a valid time reference
variable will not always define an ideal clock relative to another time
reference variable---for an example look at the discussion of constraint
$\hat{C}_1$\ below.

A useful physical clock need not be ideal: for example $\hat{U}$\ could
represent a degree of freedom that oscillates with a stable frequency like an
atomic clock. However, unless the clock is also a valid reference time, there
is, in general, no exact temporal reference frame associated with it. If we
select an algebraic reference state that satisfies the conditions laid out in
section~\ref{sec:APstates} relative to a clock $\hat{U}$\ that does not
satisfy the conditions described in section~\ref{sec:GaugeFlows}, such a state
will not retain positivity during time-evolution and, therefore, would not
result in future probabilistic predictions for measuring observables that
commute with $\hat{U}$. In this light it is interesting to consider the
interacting clock models used in~\cite{QuantumRefLocal}\ and models for clocks
experiencing an external gravitational field of different strengths employed
in~\cite{QuantumRef6,QuantumRefEquiv}. Here we are not attempting to analyze the
claims about real physical behavior of clocks subject to non-uniform
gravitational interactions, but merely point out that, according to our
algebraic analysis, whether such clocks define exact temporal quantum
reference frames sensitively depends on the interactions one introduces.

For a concrete example we will take a closer look at one of the models
in~\cite{QuantumRefLocal}, the same algebraic analysis can be applied to other
models considered there and in~\cite{QuantumRef6,QuantumRefEquiv}. The Hamiltonian
constraint in equation~(12) in~\cite{QuantumRefLocal} models two gravitationally
interacting clocks, $A$\ and $B$, a third distant clock, labeled $C$, and an
event localized in time relative to clock $A$. In our notation, the constraint
has the form
\begin{equation}\label{eq:C1general}
\hat{C}_1 = \hat{p}_A+\hat{p}_B +\hat{p}_C + \lambda_1 \hat{p}_A \hat{p}_B+ f_A(\hat{q}_A)\left( \mathbb{I}+ \lambda_2 \hat{p}_B \right)  \ .
\end{equation}
Here $f_A(\hat{q}_A)$\ is a function of $\hat{q}_A$\ sharply localized around
a (real) reading $t_A$ of clock $A$\ and multiplied by an additional operator
that couples this clock to a record-keeping subsystem, while $\lambda_i$\ are
numerical coefficients that characterize the gravitational interaction of
clocks $A$\ and $B$. (Larger $\lambda_i$ correspond to clocks being closer to
each other and interacting more strongly.) The purpose of $f_A(\hat{q}_A)$ is
to set up a ``recorded event'' that is temporally localized relative to clock
$A$. Here we will ignore its action on the record-keeping subsystem and use
the fact that $f_A(\hat{q}_A)$\ has vanishing commutators with $\hat{p}_B$,
$\hat{p}_C$, and all three configuration observables $\hat{q}_I$, $I=A, B, C$,
while $[\hat{p}_A, f_A(\hat{q}_A)] \neq 0$.

In~\cite{QuantumRefLocal} the two coupling constants are equal
$\lambda_1=\lambda_2=: \lambda$. With this choice one can re-define the
momentum of clock $A$, $\hat{p}'_A:=\hat{p}_A+f_A(\hat{q}_A)$ which retains
canonical commutators with $\hat{q}_A$\ and observables of the other two
clocks. The constraint can then be re-written
\begin{equation}\label{eq:C1simple}
\hat{C}'_1=\hat{p}'_A + \hat{p}_B+\hat{p}_C + \lambda \hat{p}'_A\hat{p}_B \ .
\end{equation}
Because $[\hat{q}_C, \hat{C}'_1] = i\hbar \mathbb{I}$, the non-interacting
distant clock $C$\ is a valid time reference variable with trivial
factorization $\hat{C}_{H;\, C}=\hat{C}'_1$\ and
$\hat{\mathcal{N}}_C=\mathbb{I}$. Furthermore, the interaction term in this
model on its own does not destroy the ideal nature of clocks $A$\ and $B$,
which can be used as time reference variables with factorization, for example
for $A$
\[
\hat{C}'_1 = \left( \mathbb{I} + \lambda \hat{p}_B \right) \left( \hat{p}'_A + \frac{\hat{p}_B +\hat{p}_C}{\mathbb{I} + \lambda \hat{p}_B} \right) : = \hat{\mathcal{N}}_A \hat{C}_{H;\, A} \ ,
\]
valid on the domain where $\left(\mathbb{I} + \lambda \hat{p}_B\right)$\ is
invertible. Since this operator commutes with $\hat{C}_{H;\, A}$, its
invertibility is preserved by time evolution. Factorization for clock $B$\ is
entirely symmetric. Note that, in this model, if e.g. $\hat{q}_C$\ defines the
temporal reference frame, the clock $\hat{q}_A$\ is no longer ideal, since its
rate
\[
[\hat{q}_A, \hat{C}_{H;\, C}] = [\hat{q}_A, \hat{C}'_1] 
= i\hbar \left( \mathbb{I} + \lambda \hat{p}_B \right) \ ,
\]
which is not of the form $\alpha i\hbar \mathbb{I}$, for
$\alpha \in \mathbb{R}$. Nevertheless, it is still a ``good'' clock since its
rate is a constant of motion
$[[\hat{q}_A, \hat{C}_{H;\, C}], \hat{C}_{H;\, C}]=0$. The same scenario
plays out if we pick another pair as reference time and a physical clock.

If $\lambda_1\neq\lambda_2$, the constraint can no longer be written in the
simple form~(\ref{eq:C1simple}). Clocks $A$\ and $C$\ remain valid time
reference variables, but clock $B$\ does not: the combination of
interaction with clock $A$, and the event-recording process spoil its behavior
in this model. The quickest way to see that the algebraic conditions for
factorization are violated is to note that $\hat{C}_1$\ is linear in
$\hat{p}_B$, so that
\[
\hat{\mathcal{N}}_B=\frac{1}{i\hbar}[\hat{q}_B, \hat{C}_1 ] = \mathbb{I} + \lambda_1 \hat{p}_A + \lambda_2 f_A(\hat{q}_A) \ ,
\]
from which it follows that
\[
[\hat{\mathcal{N}}_B, \hat{C}_1] = i\hbar (\lambda_1-\lambda_2)[\hat{p}_A,f_A(\hat{q}_A)] \neq0 \ ,
\]
violating the requirements of section~\ref{sec:GaugeFlows}. If we attempt to factorize this constraint
\[
\hat{C}_1 = \left( \mathbb{I} + \lambda_1 \hat{p}_A + \lambda_2 f_A(\hat{q}_A) \right) \left[ \hat{p}_B+ \frac{1}{\mathbb{I} + \lambda_1 \hat{p}_A + \lambda_2 f_A(\hat{q}_A)} \left( \hat{p}_A +\hat{p}_C + f_A(\hat{q}_A)\right) \right]  \ ,
\]
we end up with a non-hermitian factor on the right because $ \left( \mathbb{I} + \lambda_1 \hat{p}_A + \lambda_2 f_A(\hat{q}_A) \right)^{-1}$\ does not commute with $\left( \hat{p}_A +\hat{p}_C + f_A(\hat{q}_A)\right)$. In other words, with this minor change in the model of coupling, the ticks of clock $B$\ no longer map out a unitary evolution history of the rest of the system. 

\section{Implications for relational quantum mechanics}

Before we conclude, we would like to point out additional implications of our
algebraic discussion in a somewhat different context, given by attempts to
define (or rule out) a technical description of interacting quantum systems consistent with the stated principles of relational quantum mechanics (RQM) of
\cite{RQM,RQMRev}.  The distinction between ordinary (positive) quantum states
that freely carry correlations and constrained (almost-positive) reference
states with partially restricted correlations may be relevant to a recent
debate in this context \cite{RQM2, RQM3, RQM4, RQM5, RQM6}.

Relational approaches to quantum mechanics including RQM were inspired by the older
attempts \cite{GenHamDyn1,BergmannTime} to define meaningful observables in
quantum gravity, a theory in which the geometry of space and time is subject
to the rule of quantum mechanics. However, since space and time coordinates
are not part of the underlying phase space because they represent redundant
information, it is impossible to describe observables in a way similar to the
classical method of, say, geodesics on a space-time manifold used to set up
the local inertial frame of an observer who measures properties of a moving
object. Instead of using a time coordinate, the gist of those older proposals
was to describe evolution relationally, for instance by specifying the
position of one particle relative to the position of another particle. In this
example the second particle is used as a clock, and the first particle can be
said to evolve with respect to time as determined by the second particle.

The more recent program of RQM \cite{RQM} takes a relational view of \emph{all}
measurements because they are always performed relative to some system
describing the observer, even if gravity is not quantized. (The review
\cite{RQMRev} emphasizes the connection with quantum gravity.) Both sides of
the ongoing debate about the consistency of RQM make some use of ordinary
entangled quantum states when describing one subsystem from the perspective of
another (although the proponents of RQM prefer to de-emphasize the role of
states in favor of interactions). Such entangled states provide no obvious
relation to the description of that same subsystem from a third perspective, as, for example, given
by $\hat{q}_I$, $I=A, B, M$, in section~\ref{sec:Clocks}.
Since the choice of a reference degree of freedom for relational statements is
not unique, however, it is necessary to consider multiple perspectives and to
demonstrate some degree of invariance of physical statements with respect to
changes of relational dependencies.

If consistent transformations of reference choices exist, the overall
description of all the relevant interacting subsystems together in one
consistent setting necessarily carries a large degree of redundancy.
Different choices of reference within the same redundant setting are related by gauge
transformations, enforced by some form of quantum constraints. As a result,
and as per our discussion, any relational setting necessitates certain
restrictions on which subsystems can be simultaneously assigned positive quantum states
(i.e. given a probabilistic quantum description), and a full state for the entire
system including reference degrees of freedom can only be almost-positive.
The set of available states that can be used to construct suitable versions of
relational quantum mechanics (or counter-examples to their consistency) is
therefore limited.

\section{Discussion}

Our new notion of almost-positive algebraic states provides a consistent
embedding of the states associated with all valid time reference frames of a
system with a Hamiltonian constraint as states on the full kinematical
algebra. Such systems are algebraically different from unconstrained systems
in that the presence of the constraint and corresponding reference variables
requires a weakened form of positivity of states. Almost-positivity ensures
that the resulting evolution picture relative to the reference variable is
consistent, but results in a state that does not give probabilistic
predictions for the measurement of some kinematical observables. As we have
seen, in almost-positive states reference variables themselves do not behave
in a quantum manner: They do not fluctuate, $\Delta Z=0$, and they are
uncorrelated with system degrees of freedom, $C_{ZA}=0$. They are, in fact,
not observables, but generally require additional subsystems in order to be
measured. In a consistent treatment of quantum constrained systems, a quantum
reference frame therefore remains largely classical. In particular, time and
space in quantum gravity would be represented by reference variables that
characterize the action of constraints. If exact time and space reference
variables are found, the resulting space-time that they will map out should
retain many of its classical properties: time does not fluctuate and is not
correlated with other subsystems.

It may certainly be possible that physical clocks, defined as quantum devices
that measure time but do not represent time at a fundamental level, exhibit
fluctuations or correlations that limit our observational access to time. On
the purely mathematical level, the reference variable of one temporal
reference frame can serve as a clock in another temporal reference frame and
possess quantum fluctuations and correlations when viewed from the latter
perspective. However, our general discussion shows that correlations or
entanglement between a physical clock and some system are not necessarily
implied by the mere fact that the combined state is described with respect to
a quantum reference frame: Since any positive state on the commutant of
$\hat{Z}$ is uniquely extended to an almost-positive state of the full
algebra, the presence of a quantum reference frame does not impose any new
conditions on correlations and entanglement, other than those features that
may be implied by interactions between clock and system in the standard
way. Given the same interactions, the same correlations and entanglement would
evolve if one were to use a classical-type background time. There is then no
fundamental limitation on the accessibility of time measurements because
interactions between clock and system depend on how the clock is constructed
and can in principle be reduced by judicious choices or placements of clocks.

At the current stage of developments, mathematical consistency requirements,
rather than practical questions, seem to play a more decisive role in the
admissibility of clock-system interactions.  As we saw from the example in
section~\ref{sec:Clocks}, and a related one in the appendix, the viability of
a kinematical observable as a time reference variable is very sensitive to the
interactions one includes. A general quantum system subject to a sensible
Hamiltonian constraint may posses one or more reference times, where there may
or may not be ideal physical clocks corresponding to each choice of time; it
may also possess no valid reference times at all (and therefore also no
corresponding ideal clocks). The latter case calls for state-based
approximations, where in some states a given reference time can be used to
characterize a portion of evolution (as was done within the semiclassical
approximation in~\cite{EffTime, EffTimeLong, EffTimeCosmo}). Here too the
ability to embed all (approximate) reference time perspectives under one roof
becomes important and the algebraic construction provides a helpful framework
(see~\cite{AlgebraicFrozen}). If admissible interactions can be classified
completely for a given set of reference, clock, and system variables, they may
lead to a restricted set of possible outcomes for correlations and
entanglement between clock and system. At present, however, such a
classification is unavailable.

\section*{Acknowledgements}

This work was supported in part by NSF grant PHY-2206591.

\appendix

\section[\appendixname~\thesection]{A clock model with time dilation}

As a specific application of our results, we here outline how recent
constructions of clocks in space-time \cite{QuantumRef6,QuantumRefEquiv} fit within
the algebraic framework and the new restrictions it imposes on
deparameterizability as relational evolution.  We use the slightly simpler
constraint system of~\cite{QuantumRef6}, focusing on the example of
equation~(17) there. The system consists of two light particles $A$\ and $B$\
and a massive particle $M$\ on a $1+1$-dimensional space-time, with corresponding basic
canonical observables
$[\hat{q}^i_J, \hat{p}^K_l] = i\hbar \mathbb{I} \delta^K_J \delta ^i_l$. Here,
$i, l = 0, 1$ are space-time indices and $J, K = A, B, M$ particle indices. The
light particles also possess internal clocks
$[\hat{Z}_I, \hat{E}_J] = i\hbar \mathbb{I} \delta_{I\, J}$, with $I, J=A, B$.
(The model in~\cite{QuantumRefEquiv} includes an internal clock for the massive
particle and an additional constraint.)

There are four constraints. The first two impose the energy-momentum relations
for the light particles,
\begin{equation}\label{eq:WorldLine}
\hat{C}_I = \hat{g}_I\, \hat{p}_0^I - \hat{\omega}_I
\end{equation}
where $\hat{g}_I$\ depends on the metric at the location of the light particle
$I=A, B$, distorted from Minkowski metric by the presence of the massive
particle $M$,
\[
\hat{g}_I = \sqrt{g^{00} \left( \hat{q}^1_I-\hat{q}^1_M \right)} \ ,
\]
and $\hat{\omega}_I$\ only depends on the spatial momentum of particle $I$
\[
\hat{\omega}_I = \sqrt{m_I^2 c^2 \mathbb{I} + \left(\hat{p}_1^I\right)^2} \ .
\]
The third constraint enforces spatial translational invariance
\begin{equation}
\hat{f}^1=\hat{p}_1^A +\hat{p}_1^B +\hat{p}_1^M \ .
\end{equation}
The last is the overall Hamiltonian constraint
\begin{equation}\label{eq:HamConstraint}
\hat{f}^0 = \hat{p}_0^A +\hat{p}_0^B +\hat{p}_0^M + m_A (\hat{g}_A)^{-1} (\hat{\omega}_A)^{-1} \hat{E}_A + m_B (\hat{g}_B)^{-1} (\hat{\omega}_B)^{-1} \left(  \hat{E}_B+ \hat{\theta}_B\right) \ ,
\end{equation}
where $\hat{\theta}_B$\ is a function of the internal configuration of the
clock $\hat{Z}_B$\ that is sharply localized about a specific value $t_B$\ (in
fact~\cite{QuantumRef6} uses a delta function), multiplied by an operator that
commutes with all four constraints. (One way to construct such an operator is
to add a ``record-keeping'' subsystem that does not interact with the rest of
the system otherwise.)

While the first three constraints are manifestly hermitian and all explicitly
commute with each other, $\hat{f}^0$\ is not because $[ \hat{g}_I,
\hat{\omega}_I] \neq 0$. In~\cite{QuantumRef6} it is assumed that the value of
this commutator is small when applied to states, and that
\begin{equation}\label{Comm}
  \left[ (\hat{g}_I)^{-1}, (\hat{\omega}_I)^{-1} \right] =
  (\hat{\omega}_I)^{-1}(\hat{g}_I)^{-1}[\hat{g}_I, \hat{\omega}_I](\hat{g}_I)^{-1} (\hat{\omega}_I)^{-1} 
\end{equation}
enjoys the same property. Time dilation is therefore required to be position-independent.

Our aim here is to show that this condition is not only an approximation that
may simplify some calculations, but is in fact necessary in order to
deparameterize the constraints~(\ref{eq:WorldLine}) with respect to
$\hat{q}_I^0$ as the clock. In this model, deparameterization is achieved by
factorization
\begin{equation}\label{CI}
\hat{C}_I = \hat{g}_I \left( \hat{p}_0^I - (\hat{g}_I)^{-1}\hat{\omega}_I \right) 
\end{equation}
with $\hat{\cal N}=\hat{g}_I$ in our previous notation. If
$[ \hat{g}_I, \hat{\omega}_I] \neq 0$, the right factor in (\ref{CI}) is not hermitian and
cannot play the role of a consistent evolution generator $\hat{C}_H$. (See our
discussion in section~\ref{sec:GaugeFlows}.)

If this commutator as well as (\ref{Comm}) can be
dropped, then all constraints commute with each other and the constrained
system can be solved by removing each of them one at a time.
\begin{itemize}
\item First, we eliminate $\hat{f}^1$\ by removing $\hat{p}^M_1 = -\hat{p}_1^A
  -\hat{p}_1^B$\ and switching to relative positions
  $\hat{Q}^1_I=\hat{q}^1_I-\hat{q}^1_M$, $I=A, B$.  
\item Second, we deparameterize the constraints~(\ref{eq:WorldLine}) by
  factorization with respect to $\hat{q}_I^0$\ as clocks. Individually,
  space-time degrees of freedom of each light particle are reduced to the
  observables $\hat{q}^1_I$\ and $\hat{p}_1^I$\ in the commutant of the
  clocks. They evolve in parameter time $\tau_I$\ with respect to the
  Hamiltonian $\hat{h}_I=- (\hat{g}_I)^{-1}\hat{\omega}_I $. The
  constraint~(\ref{eq:HamConstraint}) reduces to
\[
\hat{f}^0 =  -\hat{h}_A -\hat{h}_B +\hat{p}_0^M + m_A (\hat{g}_A)^{-1} (\hat{\omega}_A)^{-1} \hat{E}_A + m_B (\hat{g}_B)^{-1} (\hat{\omega}_B)^{-1} \left(  \hat{E}_B+ \hat{\theta}_B\right) \ .
\]
\item Finally, the above constraint can be deparameterized directly by using $\hat{q}_C^0$\ as time reference, or by factorization if we instead use $\hat{Z}_A$\ or $\hat{Z}_B$. For example, in the case of $\hat{Z}_A$\ we factorize
\[
\hat{f}^0 = m_A (\hat{g}_A)^{-1} (\hat{\omega}_A)^{-1} \left[ \hat{E}_A - \frac{(\hat{\omega}_A)^2}{m_A}  + \frac{\hat{\omega}_A\hat{g}_A}{m_A} \left( \hat{p}_0^M -\hat{h}_B  + m_B (\hat{g}_B)^{-1} (\hat{\omega}_B)^{-1} \left(  \hat{E}_B+ \hat{\theta}_B\right) \right) \right] \ .
\]
This latter deparameterization, once again, requires that we ignore the
commutator $[ \hat{g}_I, \hat{\omega}_I]$, to make the third term inside the
square parentheses hermitian, and to make the left factor
$m_A (\hat{g}_A)^{-1} (\hat{\omega}_A)^{-1}$\ a constant of motion.
\end{itemize}
These constructions and our general discussion indicate that an extension of
time dilation in clock models to more relativistic settings, in which
commutators such as $[\hat{g}_I,\hat{\omega}_I]$ can no longer be ignored,
will be more challenging.


\end{document}